\documentclass[12pt,thmsa]{revtex4}
\usepackage{epsfig}

\begin{document}

\title {Negative magnetoresistance of ultra-narrow superconducting nanowires in the resistive state}

\author{K. Yu. Arutyunov} 

\address  {University of Jyv\"askyl\"a, Department of Physics, NanoScience Centre, PB 35, 40014 Jyv\"askyl\"a, Finland}

\begin{abstract}

We present a phenomenological model qualitatively explaining negative magnetoresistance in quasi-one-dimensional  superconducting channels in the resistive state. The model is %%@
based on the assumption that fluctuations of the order parameter (phase slips) are responsible for the finite effective resistance of  a narrow superconducting wire sufficiently close %%@
to the critical temperature. Each fluctuation is accompanied by an instant formation of a quasi-normal region of the order of the non-equilibrium quasiparticle relaxation length %%@
'pinned' to the core of the phase slip. The effective time-averaged voltage measured in experiment is a sum of two terms. First one is the conventional contribution linked to the rate %%@
of the fluctuations via the Josephson relation. Second term is the Ohmic contribution of this quasi-normal region. Depending on material properties of the wire, there might be a %%@
range of magnetic fields where the first term is not much affected, while the second term is effectively suppressed contributing to the experimentally observed negative %%@
magnetoresistance.

\end{abstract}


\maketitle

%  ############### end of front matter###############%

\newpage

\section{Introduction}

\label{I}

It is a text-book knowledge that shape of a superconducting phase transition 
$R(T)$ of a homogeneous sample is determined by fluctuations of the order parameter. The rounded top part
(the onset of superconductivity) is accounted for nucleation of the superconducting phase shunting the normal
current. The effect can be observed in superconducting systems irrespectedly of their dimensionality, set by the
temperature-dependent coherence length $\xi (T)$. At the bottom (low temperature) part of the transition,
dimensionality plays an essential role. As soon as at least one channel of supercurrent nucleates, in 2D and 3D systems
spots of normal phase do not contribute to zero DC resistance, and the $R(T)$ dependence has an abrupt
bottom part. Situation is qualitatively different in long quasi-1D systems with cross-section $\sigma
\leq \xi ^{2}$ where there exist only one parallel channel of a supercurrent. Fluctuations of the order parameter
destroy phase coherence in the wire and, hence, finite resistance is produced. In experiment this
phenomenon manifests itself as a rounded low temperature part of the $R(T)$ phase transition. In a very simplified
way such a behavior can be understood as for sufficiently long 1D channel of length $L\gg \xi $ there
is always a finite probability of a fluctuation to drive instantly a fraction of the wire into a normal state.
As there is only one supercurrent channel, these events cause momentary energy dissipation. The instant voltage jumps,
being integrated in time, are commonly interpreted as a finite DC resistance. If at a given measuring
current $I$ the rate of fluctuations is $\Gamma _{PS}$, then the effective resistance is given by the Josephson
relation: $R_{PS}\equiv \langle V\rangle _{PS}/I%%@
=h\Gamma _{PS}/2eI$. Two different scenarios have been developed defining
the rate of fluctuations of the order parameter (phase slips) in 1D superconducting channels: thermal (TAPS) 
\cite{TAPS} - \cite{Vinokur} and quantum (QPS) \cite{Saito} - \cite{Meidan}. The principal difference is that
in TAPS the required energy for the fluctuation is provided by the 'classical' term $\sim k_{B}T$, while in QPS
the relevant energy scale is $\sim h%%@
\Gamma _{QPS}$.

In early experiments on tin whiskers \cite{R(T)of tin whisker} it has been
shown that the shape of the $R(T)$ transition can be nicely fitted by the TAPS model \cite{TAPS}. By modern
standards the cross section of that samples $\sigma ^{1/2}\sim 0.5$ m$\mu $ is rather large, and the
corresponding width of experimentally detectable $R(T)$ transition is about $\sim 1$ mK. During the last decades the rapid
progress in nanotechnology enabled study of much narrower superconducting channels reaching sub-10 nm scales 
\cite{Graybeal} - \cite{Zgirski PRB2007}. In these ultra-narrow
nanostructures the {R(T)} transitions are much wider than the TAPS model predictions. 
The effect is associated with manifestation of the QPS mechanism. Though the
interpretation of results is still contradicting, the experiments have revealed several
unusual features. Among them - 'enhancement' of superconductivity by magnetic field, which manifests itself
either as negative magnetoressitance (nMR) or/and increase of the critical current $I_{c}$ in
magnetic field. The effect is observed in the very narrow quasi-1D superconducting channels in a resistive
state below the critical temperature $T_{c}$ \cite{Sharifi}, \cite{Zgirski PRB2007}, \cite{Bezryadin
NMR}. A trivial explanation related to Kondo mechanism can be ruled out as concentration of magnetic impurities in
these samples is negligible; the corresponding magnetic field is too small to polarize any imaginable
magnetic momentum, and additionally the onset of superconductivity is not affected by a small magnetic field where
nMR is observed. While various scenarios have been proposed \cite{Sharifi}, \cite{Zaikin}, \cite{Pesin}, \cite{Bezryadin NMR theory}, \cite{Vodolazov nMR} 
the matter is not yet settled and there is no commonly accepted explanations for the nMR phenomenon. In this
paper we present a simple phenomenological model accounting the nMR effect to formation of a
quasi-normal region 'pinned' to each fluctuation-mediated phase slip. The extension of this region is set by the quasiparticle 
relaxation length $\Lambda _{Q}$ and is effectively suppressed by a weak magnetic field giving rise to the nMR.

\section{The model}
\label{II} 

For concreteness we consider that the resistive state of a
quasi-1D superconducting channel is govern solely by the TAPS mechanism. However, we believe that qualitatively the
model should be applicable to QPS activation as well. It has been shown \cite{TAPS} that the rate of TAPS $%
\Gamma _{PS}$ at a bias DC current $I$ is:

\begin{equation}
\Gamma _{PS}(T,I)=\Omega (T,I,H)\exp \left[ -\frac{\Delta F_{0}}{k_{B}T}%
-\left( \frac{2}{3}\right) ^{1/2}\frac{I^{2}}{3\pi I_{1}I_{c}}\right] \sinh
\left( \frac{I}{2I_{1}}\right)   
%\tag{1}
\end{equation}%
, where $k_{B}$ is the Boltzman constant, $I_{1}=k_{B}T/\phi _{0}$ and $\phi
_{0}$ is the superconducting flux quantum. Condensation energy of the
miniumum superconducting domain of size $\xi \times \sigma $ is $\Delta
F_{0}\sim B_{c}^{2}\xi \sigma $, where $B_{c}(T)$ is the critical magnetic field. The
exact expression for the pre-facor $\Omega (T,I,H)$ has been defined \cite%
{TAPS}, while being not quantitatively important taking into consideration
strong exponential dependence in Eq. (1).  Utilizing the above expression
for the rate of fluctuations one can get the \textit{time-averaged}
effective voltage using the Josephson relation $\langle V\rangle
_{PS}=h\Gamma _{PS}/2e$ and the corrsponding effective resistance \cite{TAPS}%
:

\begin{equation}
R_{PS}\equiv \langle V\rangle _{PS}/I=h\Gamma _{PS}/2eI  
%\tag{2}
\end{equation}

However, one should keep in mind that the fluctuation-governed resistive
state of a 1D superconductor in reality is a dynamic process consisting of
discrete events (phase slips) repeated on average with the rate $\Gamma _{PS}
$. During each phase slip of duration $\sim h/\Delta $ inside the PS core
region $\sim \xi $ the magnitude of the order parameter goes to zero and the
phase flips by $2\pi $. It is natural to assume that each such fluctuation
gives rize to a 'quasi-normal' region where a conversion from
non-equilibrium quasiparticles to equilibrium Cooper pairs takes place. 
The charge imbalance is 'pinned' to each phase
slip and decays in time and space. The energy dissipation inside this
quasi-normal region  should be taken into account to obtain the total
effective resistance. The main assumption of the present model is that the
processes describing the non-equilibrium state at static normal metal -
superconductor interfaces \cite{Tinkham book} and/or current induced resistive state in 1D superconductors \cite{Tidecks} are applicable to the fluctuation-governed 1D
superconductivity.

The relaxation of the charge imbalance $Q$ in a 1-D superconducting channel can
be described by the differential equation \cite{Kadin}: 
\begin{equation}
D\tau _{Q}\bigtriangledown ^{2}Q=\tau _{0}\tau _{E}\ddot{Q}+(\tau _{0}+\tau
_{E})\dot{Q}+Q  
%\tag{3}
\end{equation}%
, where $\tau _{E}$ is the inelastic electron-phonon collision time, $\tau
_{0}=2k_{B}T_{c}\hbar /\pi \Delta ^{2}$ is the supercurrent response time,
and $D=\frac{1}{3}v_{F}l$ is the diffusion coefficient, $v_{F}$ being the Fermi
velocity and $l$ the electron mean-free path. In the most general case
Eq.(3) describes damped, dispersive waves of charge imbalance.  However, for
majority of practical applications the low-frequncy limit is applicable: $%
\tau _{0}\ll \tau _{E}\ll 1/\Gamma _{PS}$. In this case the charge imbalance
decays on the length $\Lambda _{Q}=(D\tau _{Q})^{1/2}$.  The characteristic
relaxation time $\tau _{Q}$ is given by \cite{Schmid-Shoen}: 

\begin{equation}
\frac{1}{\tau _{Q}}=\frac{\pi \Delta }{4k_{B}T_{c}}\frac{1}{\tau _{E}}\left[
1+\frac{2\tau _{E}}{\tau _{S}}\right] ^{1/2}  
%\tag{4}
\end{equation}

, where $\tau _{S}$ is the pair-breaking time due to supercurrent or
magnetic field. In the most general case the gap parameter is a function
of temperature, current and magnetic field $\Delta =\Delta (T,I,H)$. The
extra Ohmic contribution associated with dissipation from both sides of the
PS core is:

\begin{equation}
R_{Q}=\rho _{Q}\left( \frac{2\Lambda _{Q}}{\sigma }\right) (\tau _{0}\Gamma
_{PS})  
%\tag{5}
\end{equation}

, where $\rho _{Q}$ is the effective resistivity of the quasi-normal region.
In the simplest approach $\rho _{Q}$ can be considered of the order of the
normal state resitivity of the material. The last term in Eq. (5) is the
statistical wieght accountintig for the repetition of PSs with the average
rate $\Gamma _{PS}$ each of duration $\tau _{0}$. To obtain the total
effective resistance of a 1D channel in the resistive state one should take
a sum of Eq. (2) and Eq. (5).

\begin{figure}[h]
\begin{center}
\includegraphics  [width=1 \textwidth] {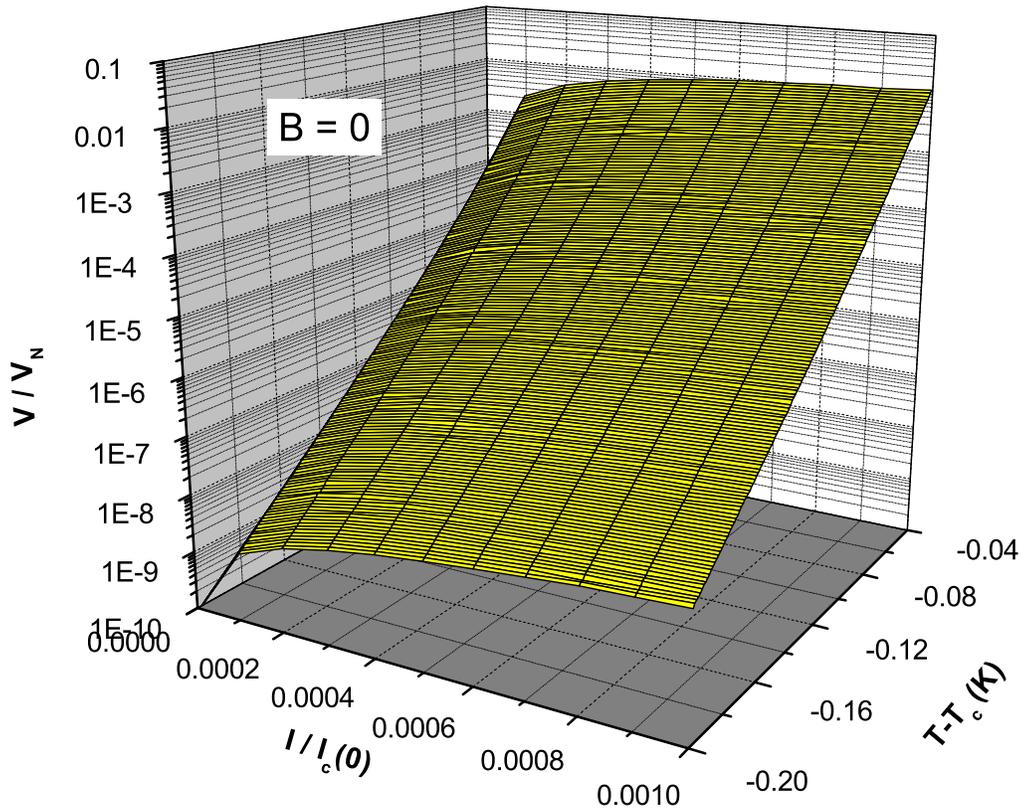}
\end{center}
\caption{Effective time-averaged voltage vs. supercurrent $I$ and temperature $T$ in zero magnetic field for 20 nm $\times$ 20 nm $\times$ 10 $\mu$m aluminum nanowire. Current %%@
is normalized by its zero-temperature value $I_{c}(0)$.}
\label{curent}
\end{figure}

The pair-breaking due to supercurrent \cite{Maki}, \cite{Lemberger} is 
 $1/\tau _S^I=D(p_S/\hbar )^2/2$, where $p_S$ is the supercurrent
momentum. Using familiar expressions \cite{Tinkham book} the $\tau _S^I$ can be
re-written in a more convenient form: 

\begin{equation}
\frac 1{\tau _S^I}=\frac D2\left[ \frac I{3\sqrt{3}I_c(T)\xi (T)}\right] ^2 
%\tag{6}
\end{equation}

As it is clearly seen from Eq. (6) and Eq. (5) the current reduces the quasiparticle relaxation time and, hence, decreases the charge imbalance relaxation length. However, the %%@
supercurrent $I$ always brings the rate of thermal fluctuations $\Gamma _{PS}$ to higher values: Eq. (1). Note that formally one has to consider the variation of the gap $\Delta %%@
=\Delta (I)$ with the applied current. In the present paper we neglect this small $\Delta (I)$ deviation which plays no significant quantitative role for calculations.  The resulting %%@
time-averaged voltage always increases with current (Fig. 1). 

The impact of magnetic field on the charge imbalance relaxation is qualitatively similar to supercurrent: magnetic field decreases the pair breaking time \cite{Kadin-Smith-Tinkham}:

\begin{equation}
\frac{1}{\tau _{S}^H} = \frac{1.76k_BT_c}{\hbar} \frac{H^2}{H_{c}^{\parallel }(0)^{2}}  
%\tag{7}
\end{equation}

where $H_{c}^{\parallel }(0)$ is the zero-temperature parallel critical
field. Contrary to the case of pair-breaking currents the variation of the
gap with magnetic field $\Delta (T,H)= \Delta (T) [1-H^2/H_c^{\parallel }(T)^2]^{1/2}$ 
plays quantitatively significant role. Slightly affecting the $\Gamma _{PS}$ due to variation of the gap $\Delta (H)$, magnetic field noticeably shifts $\tau _{Q}(H)$ to shorter values. %%@
For each particular set of parameters (material, wire diameter, temperature) there is a range of magnetic fields well below the critical values, where the quasiparticle charge %%@
imbalance length (= quasi-normal region) is reduced, while the rate of fluctuations is not affected much. The interplay between these two mechanisms gives rise to the negative %%@
magnetoresistance (Fig. 2). Note that the effect should be present in all superconducting systems falling into the 1D limit $\sigma ^{1/2}\leq \xi$. However, the experimentally %%@
observable finite width of the superconducting transition exponentially depends on the wire cross section $\sigma$ (Eq. (1)). Hence, observation of the negative magnetoresistance %%@
phenomenon requires ultra-narrow superconducting channels, and has been observed so far only in sub-100 nm nanostructures \cite{Sharifi}, \cite{Zgirski PRB2007}, %%@
\cite{Bezryadin NMR}.

\begin{figure}[h]
\begin{center}
\includegraphics[width=1 \textwidth]{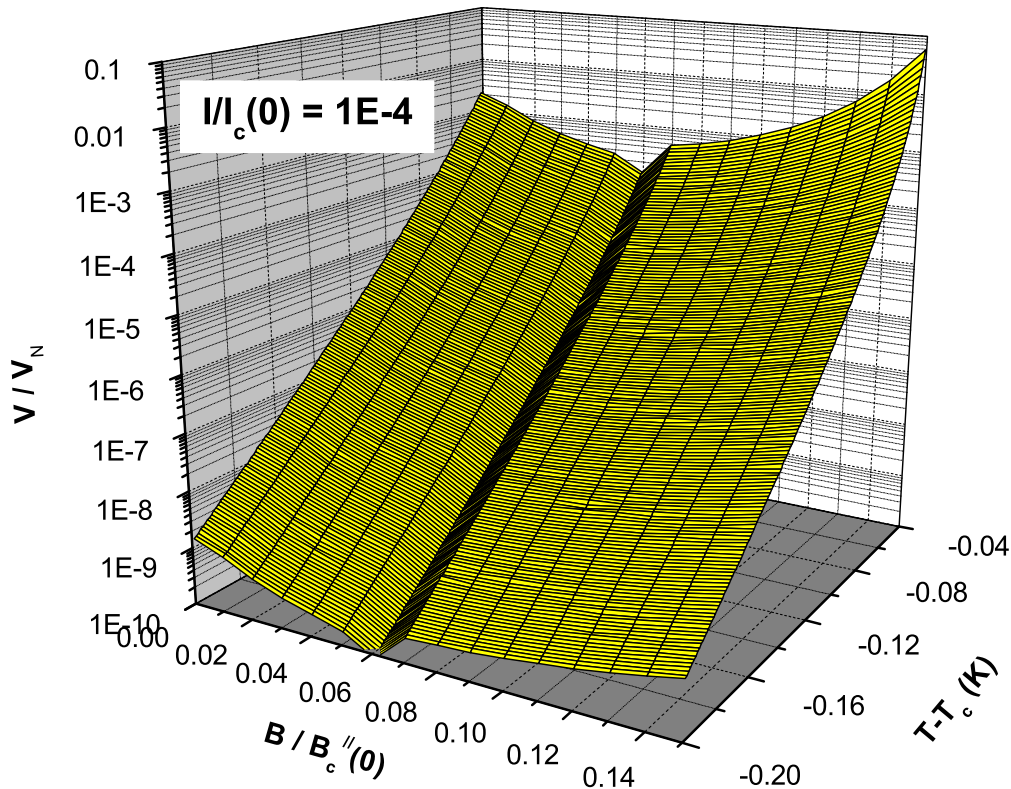}
\end{center}
\caption{Effective time-averaged voltage vs. magnetic field $B$ and temperature $T$ at constant DC measuring  current $I$ normalized by its zero-temperature value $I_{c}(0)$ for 20 %%@
nm $\times$ 20 nm $\times$ 10 $\mu$m aluminum nanowire. Magnetic fields are normalized by the zero-temperature parallel critical magnetic field $B_{c}^{\parallel }(0)$. One can %%@
clearly see the negative magnetoresistance developing at small fields.}
\label{magnetic}
\end{figure}

\newpage
\section{Conclusions}
\label{III} 

In conclusion we have presented a simple phenomenological model explaining the negative magnetoresistance observed in ultra-narrow quasi-1D superconducting nanowires in the %%@
resistive state. The effect originates from competition of two mechanisms : thermodynamic fluctuations of the order parameter and quasiparticle charge imbalance, which %%@
accompanies each phase slip event. First process provides conventional positive magnetoresistance, while the second mechanism gives the negative contribution. The whole %%@
concept of thermally activated phase slips is applicable sufficiently close to the critical temperature $T_c$ where the utilized formalism of the quasiparticle charge imbalance is %%@
valid. Experimentally the nMR effect is observed in superconducting nanowires with very broad superconducting transitions $R(T)$ which can be eventually accounted for a %%@
non-thermal PS activation \cite{Sharifi}, \cite{Zgirski PRB2007}, \cite{Bezryadin NMR}. We believe that  the proposed explanation of the nMR phenomenon should be qualitatively %%@
valid also in case of the quantum phase slip (QPS) mechanism. However, in this case the extrapolation of the quasiparticle charge imbalance concept down to temperatures well %%@
below the critical one requires further justification. It might happen that in this limit the quasiparticle relaxation length $\Lambda _{Q}$ should be substituted by the other relevant %%@
scale (e.g. coherence length) \cite{Klapwijk low T NS}.

%%%%%%%%%%%%%%%%%%%%%%%%%%%%%%%%%%%%%%%%

\end{document}